\documentclass[journal]{IEEEtran}
\usepackage{amsmath, amssymb, amsthm, mathtools, relsize, paralist, hyperref}
\usepackage[T1]{fontenc}

\newtheorem{theorem}{Theorem}
\newtheorem{lemma}[theorem]{Lemma}
\newtheorem{remark}{Remark}
\newtheorem{example}{Example}

\newcommand{ \C }{ \mathcal C }

\newcommand{ \Z }{ \mathbb{Z} }
\newcommand{ \N }{ \mathbb{N} }
\newcommand{ \myx }{ \bs{x} }
\newcommand{ \myxt }{ \bs{\tilde{x}} }
\newcommand{ \myy }{ \bs{y} }
\newcommand{ \myyt }{ \bs{\tilde{y}} }

\newcommand{ \bs }[1]{ \boldsymbol{#1} }

\newcommand{ \weight }{ \operatorname{wt}_{\textsc{h}} }
\newcommand{ \myqed }{ \hfill $\blacktriangle$ }

\newcommand{ \defeq }{ \coloneqq }
\newcommand{ \defeqr }{ \eqqcolon }

\begin{document}

\title{\LARGE Asymptotically Optimal Codes Correcting Fixed-Length Duplication Errors in DNA Storage Systems}

\author{
        Mladen~Kova\v{c}evi\'c and Vincent Y. F. Tan%
\thanks{
        This work was supported by the Singapore Ministry of Education (grant no. R-263-000-C83-112).
				M. Kova\v{c}evi\'{c} was also partially supported by the European Commission (H2020 Antares project, ref. no. 739570).

        M. Kova\v{c}evi\'{c} was with the Department of Electrical \& Computer Engineering,
        National University of Singapore, Singapore 117583.
				He is now with the BioSense Institute, University of Novi Sad,
				Dr Zorana \DJ in\dj i\'{c}a 1, 21000 Novi Sad, Serbia
				(email: kmladen@uns.ac.rs).
				
				V. Y. F. Tan is with the Department of Electrical \& Computer Engineering,
        National University of Singapore, Singapore 117583, and the Department of Mathematics,
				National University of Singapore, Singapore 119076
				(email: vtan@nus.edu.sg).}%
}%


\maketitle

\begin{abstract}
A (tandem) duplication of length $ \bs{k} $ is an insertion of an exact copy of
a substring of length $ \bs{k} $ next to its original position.
This and related types of impairments are of relevance in modeling communication
in the presence of synchronization errors, as well as in several information
storage applications.
We demonstrate that Levenshtein's construction of binary codes correcting
insertions of zeros is, with minor modifications, applicable also to channels
with arbitrary alphabets and with \emph{duplication} errors of arbitrary
(but fixed) length $ \bs{k} $.
Furthermore, we derive bounds on the cardinality of optimal $ \bs{q} $-ary codes
correcting up to $ \bs{t} $ duplications of length $ \bs{k} $, and establish the
following corollaries in the asymptotic regime of growing block-length:
\begin{inparaenum}
\item[1)]
the presented family of codes is optimal for every $ \bs{q, t, k}$, in the
sense of the asymptotic scaling of code redundancy;
\item[2)]
the upper bound, when specialized to $ \bs{q=2} $, $ \bs{k=1} $, improves
upon Levenshtein's bound for every $ \bs{t\geq3} $;
\item[3)]
the bounds coincide for $ \bs{t=1} $, thus yielding the exact asymptotic
behavior of the size of optimal single-duplication-correcting codes.
\end{inparaenum}
\end{abstract}%

\begin{IEEEkeywords}
Tandem duplication, sticky insertion, deletions of zeros, repetition error,
synchronization error, bounds on codes, Sidon set, magnetic storage, DNA storage.%
\end{IEEEkeywords}

\vspace{-1mm}
\section{Introduction and Preliminaries}
\label{sec:intro}

\IEEEPARstart{T}{he emerging} technology of DNA data storage \cite{yazdi},
apart from having a multitude of applications, poses interesting new challenges
to the traditional lines of research in information theory and error control
coding.
In particular, several channel models arise in this context that are typically
not encountered in more conventional data transmission and storage systems.
Motivated by one such model that was introduced recently, we address in this
letter the problem of error correction in channels where the only impairments
are duplications of substrings in the transmitted string of symbols.
Although the main motivating examples are channels with binary or quaternary
alphabets, in the interest of generality we will in fact study channels with
arbitrary alphabets.
In the following two subsections we describe precisely the channel model we
have in mind and our contributions.

\vspace{-2mm}
\subsection{The Channel Model}
\label{sec:model}

Throughout this letter, $ \Z $ denotes the integers, $ \N $ the positive
integers, and $ \Z_q \defeq \Z/(q\Z) $ the integers modulo $ q $.

We assume that the channel alphabet, both input and output, is $ \Z_q $.
The channel inputs are strings of length $ n $ over $ \Z_q $, i.e., \newpage\noindent
elements of $ \Z_q^n $.
The channel acts on the transmitted strings by introducing multiple duplication
errors of length $ k $ in succession, where a duplication of length $ k $ is
defined as an insertion of an exact copy of a substring of length $ k $ next
to its original position; see Example \ref{exmpl} for an illustration.
We refer to this channel as the $ k $-duplication channel.

\begin{example}
\label{exmpl}
\textnormal{
Consider the following input string $ \myxt \in \Z_3^{10} $ and the corresponding
output string $ \myyt $ obtained after the channel has introduced several duplication
errors of length $ k = 3 $:
\begin{equation}
\label{eq:xy}
\begin{aligned}
  \myxt   \ = \ \  &0 \ 1 \ 1 \ 2 \ 0 \ 2 \ 1 \ 0 \ 0 \ 2  \\
	        \hookrightarrow   \ &0 \ 1 \ 1 \ 2 \ \underline{1 \ 1 \ 2} \ 0 \ 2 \ 1 \ 0 \ 0 \ 2  \\
	        \hookrightarrow   \ &0 \ 1 \ 1 \ 2 \ 1 \ 1 \ 2 \ 0 \ \underline{1 \ 2 \ 0} \ 2 \ 1 \ 0 \ 0 \ 2  \\
          \hookrightarrow   \ &0 \ 1 \ 1 \ 2 \ 1 \ 1 \ 2 \ 0 \ 1 \ 2 \ 0 \ 2 \ 1 \ 0 \ 0 \ \underline{1 \ 0 \ 0} \ 2  \ = \  \myyt .
\end{aligned}
\end{equation}
The inserted substrings at each step are underlined.
The total number of duplications that occurred in the channel is $ 3 $.
\myqed
}
\end{example}

By using the transformation $ \phi_k : \Z_q^n \to \Z_q^n $,
$ \myxt \mapsto \myx $, defined by $ x_i = \tilde{x}_i - \tilde{x}_{i-k} $,
$ 1 \leq i \leq n $, where subtraction is performed modulo $ q $ and it is
understood that $ \tilde{x}_i = 0 $ for $ i \leq 0 $, one can show that
duplication errors of length $ k $ are essentially equivalent to insertions
of blocks of $ k $ zeros, denoted $ 0^k $ \cite{jain}.
For example, for the strings in \eqref{eq:xy} and $ k = 3 $ we would have:
\begin{equation}
\label{eq:xy0}
\begin{aligned}
  \myx  \ &= \ 0 \ 1 \ 1 \ 2 \ 2 \ 1 \ 2 \ 0 \ 1 \ 1  \\
  \myy  \ &= \ 0 \ 1 \ 1 \ 2 \ \underline{0 \ 0 \ 0} \ 2 \ \underline{0 \ 0 \ 0} \ 1 \ 2 \ 0 \ 1 \ \underline{0 \ 0 \ 0} \ 1 .
\end{aligned}
\end{equation}
In particular, if a code $ \C \subseteq \Z_q^n $ can correct $ t $
insertions of blocks $ 0^k $, then $ \tilde{\C} = \phi_k^{-1}(\C) $ can correct
$ t $ duplications of length $ k $; furthermore, since $ \phi_k $ is a bijection,
we have $ |\C| = |\tilde{\C}| $.
For convenience, we will focus in the sequel on the\linebreak
$ 0^k $-insertion channel---the channel with insertions of blocks $ 0^k $ as the
only type of noise.
Due to the above-described equivalence, our main results can easily be translated
to the corresponding results for the $ k $-duplication channel:
\begin{inparaenum}
\item[(1)]
asymptotic bounds on codes for the $ 0^k $-insertion channel are automatically
valid for the $ k $-duplication channel as well, and
\item[(2)]
a construction of codes for the $ k $-duplication channel can be obtained from
a construction of codes for the $ 0^k $-insertion channel by applying the
transformation $ \phi_k^{-1} $ on the latter.
\end{inparaenum}

\vspace{-2mm}
\subsection{Previous Work and Main Results}

The binary channel with insertions of zeros was first studied in \cite{levenshtein},
where a construction of codes correcting $ t $ such errors was described and
bounds on the cardinality of optimal codes derived.
As mentioned in the previous subsection, these results are applicable to channels
with duplication errors of length $ k = 1 $ as well.
Different constructions of codes for the binary $ 1 $-duplication channel were
subsequently given in \cite{dolecek+anantharam, mahdavifar+vardy}.

A more general model, that is also studied here, with arbitrary alphabets and
duplications of length $ k $ was introduced in \cite{jain}.
In that work, in particular, optimal codes correcting \emph{all} patterns of
duplications of length $ k $ were found ($ t = \infty $).
It was also shown in \cite{jain} that optimal codes correcting $ t \in \N $
duplications of length $ k $ can be obtained from optimal codes in the $ \ell_1 $
metric.
However, constructions of optimal codes in the $ \ell_1 $ metric for general
parameters are not known at this point, and hence no estimate of the cardinality
of the resulting duplication-correcting codes was given in \cite{jain}. 
An explicit construction of codes for the special case $ t = 1 $ was recently
given in \cite{lenz}.

Our contributions can be summarized as follows:
\begin{itemize}
\item
We show that $ q $-ary codes correcting $ t $ insertions of blocks $ 0^k $
can be constructed from Sidon sets, a notion borrowed from additive combinatorics
(Theorem \ref{thm:construction}).
\item
We derive bounds on the cardinality of optimal codes of length $ n \to \infty $
correcting $ t $ insertions of blocks $ 0^k $ (Theorem \ref{thm:bounds}).
In particular, we obtain the exact asymptotic behavior of the size of optimal
single-duplication-correcting codes ($ t = 1 $), for arbitrary $ q, k $.
\item
Specializing the bounds to $ q = 2, k = 1 $, we obtain an improvement over the
best known upper bound from \cite{levenshtein} (Remark \ref{rem:q2k1}).
\end{itemize}

While this paper was under review, another work appeared \cite{lenz2} addressing
very similar problems---constructions and bounds on $ q $-ary codes correcting
$ t $ duplications of length $ k $.
The asymptotic lower bounds obtained here and in \cite{lenz2} are
the same, whereas our upper bound is strictly better than the one in \cite{lenz2},
for every $ q, k, t $.

Apart from error correction, various other problems concerning duplications
in strings were studied in the literature;
see, e.g., the references in \cite{jain, yehezkeally+schwartz}.

\vspace{-2mm}
\section{Codes Correcting Insertions and Deletions of Blocks of Zeros}
\label{sec:codes}

\subsection{General Properties}
\label{sec:properties}

The $ 0^k $-insertion channel, by its definition, affects only the lengths of
runs of zeros in the transmitted strings, it does not alter the non-zero symbols.
In particular, the Hamming weight of the transmitted string is always preserved.
This fact simplifies the analysis considerably and enables one to focus on
studying constant-weight codes without loss of generality.

We say that a code $ \C \in \Z_q^n $ can correct $ t $ insertions (resp.\ deletions)
of blocks $ 0^k $ if every codeword $ \myx \in \C $ can be reconstructed
uniquely after inserting (resp.\ deleting) up to $ t $ blocks $ 0^k $.
We say that $ \C \in \Z_q^n $ can correct $ t $ insertions \emph{and} deletions
of blocks $ 0^k $ if every codeword $ \myx \in \C $ can be reconstructed
uniquely after inserting $ t_{\textnormal{ins}} $ and deleting $ t_{\textnormal{del}} $
blocks $ 0^k $, for any $ t_{\textnormal{ins}}, t_{\textnormal{del}} $ with
$ t_{\textnormal{ins}} + t_{\textnormal{del}} \leq t $.
The following claim is a straightforward generalization of
\cite[Lem. 1]{levenshtein} to arbitrary $ q, k $, so the proof is omitted.

\begin{lemma}
\label{thm:indel}
The following statements are equivalent for every $ q, n, t, k \in \N $,
$ q \geq 2 $, and every code $ \C \subseteq \Z_q^n $:
\begin{itemize}
\item
$ \C $ can correct $ t $ insertions of blocks $ 0^k $.
\item
$ \C $ can correct $ t $ deletions of blocks $ 0^k $.
\item
$ \C $ can correct $ t $ insertions and deletions of blocks $ 0^k $.
\hfill\IEEEQED
\end{itemize}
\end{lemma}

The third point of Lemma \ref{thm:indel}, in particular, will be used in the
proof of Theorem \ref{thm:bounds} to optimize the upper bound on the cardinality
of codes correcting insertions of blocks $ 0^k $.

\subsection{Construction}
\label{sec:construction}

Let $ G $ be a finite Abelian group, written additively.
A set $ B = \{ b_1, \ldots, b_{w} \} \subseteq G $ is said to be a
\emph{Sidon set} of order $ t $ (or $ B_t $ set) if the sums
$ b_{i_1} + \cdots + b_{i_u} $ have different values for every choice of
$ u \in \{0, 1, \ldots, t\} $ and $ 1 \leq i_1 \leq \cdots \leq i_u \leq w $.
Put another way, the sums $ \sum_{i=1}^w u_i b_i $ are required to be different
for all $ u_1, \ldots, u_w \in \Z $ with $ u_i \geq 0 $, $ \sum_{i=1}^w u_i \leq t $
(here $ u_i b_i $ denotes the sum of $ u_i $ copies of the element $ b_i \in G $).
These and related objects have been studied quite extensively in combinatorics
and additive number theory; see \cite{obryant} for references.
We next describe a code construction based on the notion of Sidon sets.
The construction is a generalization of the one given in \cite{levenshtein}%
\footnote{Similar constructions of codes based on Sidon sets appear in
various contexts in coding theory;
see, e.g., \cite{derksen, graham+sloane, klove, kovacevic+tan}.
The algebraic version of the construction given here and in the mentioned works
can also be stated geometrically using the language of lattices; see \cite{kovacevic+tan, kovacevic}.}
for $ q = 2 $, $ k = 1 $.

Let $ \weight(\myx) $ denote the Hamming weight of the string $ \myx \in \Z_q^n $.
Let also $ r_i(\myx) $ denote the length of the $ i $'th run of zeros in $ \myx $.
In other words, if $ \weight(\myx) = w $, we have
$ \myx = 0^{r_0(\myx)} \alpha_1 0^{r_1(\myx)} \alpha_2 \cdots 0^{r_{w-1}(\myx)} \alpha_w 0^{r_{w}(\myx)} $,
where $ \alpha_i \in \Z_q \!\setminus\! \{0\} $.%

\begin{theorem}
\label{thm:construction}
Fix $ q, n, w, t, k \in \N $, $ q \geq 2 $, an Abelian group $ G $, a subset
$ B = \{b_1, \ldots, b_w\} \subseteq G $, an element $ b \in G $, and
define the code:
\begin{equation}
\label{eq:construction}
  \left\{  \myx \in \Z_q^n  \, : \,  \weight(\myx) = w, \;
		        \sum_{i=1}^{w}  \Big\lfloor \frac{r_i(\myx)}{k} \Big\rfloor  b_i = b
  \right\} .
\end{equation}
If $ B $ is a Sidon set of order $ t $, then the code \eqref{eq:construction}
can correct $ t $ insertions of blocks $ 0^k $.
\end{theorem}
\begin{IEEEproof}
Let $ \myx $ be the transmitted codeword and suppose that, after $ u $
insertions of blocks $ 0^k $ in the channel, the string $ \myy $ was produced
at the output.
If $ u_i $ blocks $ 0^k $ were inserted in the $ i $'th run of zeros in $ \myx $,
$ i = 0, 1, \ldots, w $, then $ r_i(\myy) - r_i(\myx) = u_i k $ and
$ \sum_{i=0}^{w} u_i = u $, where $ w = \weight(\myx) = \weight(\myy) $.
Given $ \myy $, the receiver computes the following check-sum:
\begin{equation}
\begin{aligned}
\label{eq:check}
   \sum_{i=1}^{w}  \Big\lfloor \frac{r_i(\myy)}{k} \Big\rfloor  b_i
	  =  \sum_{i=1}^{w}  \left( \Big\lfloor \frac{r_i(\myx)}{k} \Big\rfloor + u_i \right)  b_i
	  =  b  +  \sum_{i=1}^{w}  u_i b_i ,
\end{aligned}
\end{equation}
and also infers the total number of insertions $ u $ from the length of $ \myy $.
Since $ B $ is a Sidon set of order $ t $, the check-sums
$ b + \sum_{i=1}^{w} u_i b_i $ are different for all $ u_1, \ldots, u_w $
satisfying
$ u_i \geq 0 $, $ \sum_{i=1}^{w} u_i \leq  t $.
Therefore, given $ \myy $ and assuming that $ u \leq t $, the decoder can
uniquely recover the pattern of insertions $ u_0, u_1, \ldots, u_w $ by computing
\eqref{eq:check}, inferring $ u_1, \ldots, u_w $ from the result, and
concluding that $ u_0 = u - \sum_{i=1}^{w} u_i $.
\end{IEEEproof}

\vspace{2mm}
Note that the construction \eqref{eq:construction} is not explicit.
For it to be made ``practical'', one would need to describe efficient constructions
of Sidon sets, optimal ways of choosing the element $ b $, and explicit mappings
of information sequences to codewords.
Describing explicit and efficient constructions for this and related channel
models is an important problem that we shall have to leave for future investigation.

\subsection{Bounds}
\label{sec:bounds}

The following notation is used in the rest of this section:
given two non-negative real sequences $ (a_n) $ and $ (b_n) $,
$ a_n \sim b_n $ stands for $ \lim_{n \to \infty} \frac{a_n}{b_n} = 1 $, 
$ a_n \lesssim b_n $ for $ \limsup_{n \to \infty} \frac{a_n}{b_n} \leq 1 $, and
$ a_n = o(b_n) $ for $ \lim_{n \to \infty} \frac{a_n}{b_n} = 0 $.
The base-$ 2 $ logarithm is denoted by $ \log $.

We first give one auxiliary result that will be needed in the derivation of
the bounds in Theorem \ref{thm:bounds}.
Informally, it states that the ``typical'' values of the Hamming weight and
the number of runs of zeros of length $ \geq k $ in $ q $-ary strings of length
$ n \to \infty $ are $ \frac{q-1}{q} n $ and $ \frac{q-1}{q^{k+1}} n $, respectively.
To state the lemma precisely, let us denote by $ S_q^{\mathsmaller{(\geq k)}}(n,w,m) $
the number of $ q $-ary strings of length $ n $, Hamming weight $ w $, and having
exactly $ m $ runs of zeros of length $ \geq k $.

\begin{lemma}
\label{thm:typical}
Fix $ q, t, k \in \N $, $ q \geq 2 $, and define
$ \omega_q \defeq (q-1)/q$ and
$ \mu_{q,k} \defeq \omega_q (1 - \omega_q)^k = (q - 1) / q^{k+1} $.
There exists a sub-linear function%
\footnote{The function $ f $ in general depends on the constants $ q, k $ as well;
this is suppressed for notational simplicity.}
$ f(n) = o(n) $ such that, for all $ n \geq 1 $,
\begin{equation}
\label{eq:typical}
 q^n  -  \sum_{\substack{w, m \, : \, |w - \omega_q n| \leq f(n) , \\ \hskip 9mm |m - \mu_{q,k} n| \leq f(n)}} 
  S_q^{(\geq k)}(n,w,m)   \;<\;   \frac{q^n}{n^{\log n}} .
\end{equation}
\end{lemma}
\begin{IEEEproof}
The analysis parallels that in \cite[Sec. II.B]{kovacevic}, the main difference
being that the alphabet is $ q $-ary in our case, so we only give an outline.
Denote by $ S^{(j)}_q(n,w,\ell) $ the number of $ q $-ary strings of length $ n $,
Hamming weight $ w $, and having exactly $ \ell $ runs of zeros of length $ j $.
In the asymptotic regime $ n \to \infty $, $ w \sim \omega n $, $ \ell \sim \lambda n $,
for fixed $ \omega \in (0,1) $, $ \lambda \in (0, \omega) $, this quantity grows
exponentially with the exponent \cite{kovacevic}
\begin{equation}
\label{eq:exp}
\begin{aligned}
  &\lim_{n \to \infty} \frac{1}{n} \log S_q^{(j)}(n, \omega n, \lambda n) = \\
  &\omega \log(q-1) + \omega H\Big(\frac{\lambda}{\omega}\Big) +
   (\omega - \lambda) \log \sum_{\substack{i = 1\\i\neq j}}^\infty \rho_{\omega,\lambda}^{i-\frac{1-\lambda(j+1)}{\omega - \lambda}} ,
\end{aligned}
\end{equation}
where $ H(\cdot) $ is the binary entropy function, and $ \rho_{\omega,\lambda} $
is the unique positive solution to the equation:
\begin{equation}
  \sum_{\substack{i = 1\\i\neq j}}^\infty \left( i-\frac{1-\lambda(j+1)}{\omega - \lambda} \right) z^{i} = 0 .
\end{equation}
Now, since the total number of $ q $-ary strings of length $ n $ is $ q^n $,
and since there are only linearly (in $ n $) many possible weights $ w $ and
numbers of runs $ \ell $, there must exist values of $ \omega $ and $ \lambda $
for which the right-hand side of \eqref{eq:exp} (the exponent) equals $ \log q $.
Differentiating this exponent with respect to $ \omega $ and $ \lambda $, one finds
that it is \emph{uniquely} maximized for $ \omega = \omega_q = \frac{q-1}{q} $ and
$ \lambda = \omega_q^2 (1 - \omega_q)^j \defeqr \lambda_{q,j} $.
This implies that, for any given $ \epsilon > 0 $, if we exclude the strings of
weight $ w \in \big( (\omega_q - \epsilon)n , (\omega_q + \epsilon)n \big) $ having
$ \ell \in\linebreak \big( (\lambda_{q,j} - \epsilon)n , (\lambda_{q,j} + \epsilon)n \big) $
runs of zeros of length $ j $, the number of the remaining strings is
exponential with an exponent \emph{strictly smaller} than $ \log q $.
In other words, for every $ \epsilon > 0 $ there exists a (sufficiently small)
$ \delta(\epsilon) > 0 $ such that, as $ n \to \infty $,
\begin{equation}
\label{eq:typical1}
  q^n  -  \sum_{\substack{w, \ell \,:\, |w - \omega_q n| \leq \epsilon n , \\ \hskip 6mm
                                          |\ell - \lambda_{q,j} n| \leq \epsilon n}}
     S_q^{(j)}(n,w,\ell)   \,\lesssim\,   q^{(1 - \delta(\epsilon))n} .
\end{equation}
This further implies that, for every $ \epsilon > 0 $ and large enough $ n $,
\begin{equation}
\label{eq:typical2}
  q^n  -  \sum_{\substack{w, \ell \,:\, |w - \omega_q n| \leq \epsilon n , \\ \hskip 6mm
                                          |\ell - \lambda_{q,j} n| \leq \epsilon n}}
     S_q^{(j)}(n,w,\ell)   \,<\,   \frac{q^n}{n^{\log n}} .
\end{equation}
Let $ n_0(\epsilon) $ be the smallest positive integer such that \eqref{eq:typical2}
holds for all $ n \geq n_0(\epsilon) $.
Take an arbitrary sequence $ (\epsilon_i) $ satisfying
$ 1 = \epsilon_0 > \epsilon_1 > \epsilon_2 > \ldots $ and $ \lim_{i \to \infty} \epsilon_i = 0 $,
and define the function:
\begin{equation}
\label{eq:f}
  f'(n)  \defeq  \epsilon_i n ,  \qquad   n_0(\epsilon_i) \leq n < n_0(\epsilon_{i+1}) .
\end{equation}
Clearly, $ f'(n) = o(n) $.
Furthermore, from \eqref{eq:typical2} and \eqref{eq:f} we conclude that,
for all $ n \geq n_0(1) = 1 $,
\begin{equation}
\label{eq:typical3}
  q^n  -  \sum_{\substack{w, \ell \,:\, |w - \omega_q n| \leq f'\!(n) , \\ \hskip 6mm
                                         |\ell - \lambda_{q,j} n| \leq f'\!(n)}}
    S_q^{(j)}(n,w,\ell)   \,<\,   \frac{q^n}{n^{\log n}} ,
\end{equation}
which essentially completes the proof.
It is now not difficult to conclude that the relation \eqref{eq:typical} holds
as well (with a possibly different sub-linear function, $ f $).
The typical value of the number of runs of length $ \geq k $ is obtained
simply by adding up the typical values of the numbers of runs of length $ j $:
$ \sum_{j = k}^\infty \lambda_{q,j} = \omega_q (1 - \omega_q)^k = \mu_{q,k} $.
\end{IEEEproof}

\vspace{2mm}
It follows from the above proof that Lemma \ref{thm:typical} continues to hold
if $ n^{\log n} $ is replaced with an arbitrary sub-exponential function, but
this choice is sufficient for our purposes.
In particular, since $ \frac{q^n}{n^{\log n}} = o(\frac{q^n}{n^t}) $ for any
fixed $ t $, Lemma \ref{thm:typical} will enable us to disregard the non-typical
input strings in the asymptotic analysis of the size of optimal codes.

Let $ M_q(n; t; k) $ denote the size of an optimal code in $ \Z_q^n $ correcting
$ t $ insertions of blocks $ 0^k $ (or, equivalently, $ t $ insertions and deletions
of blocks $ 0^k $; see Lemma \ref{thm:indel}), and $ M_q(n, w; t; k) $ the size
of an optimal constant-weight code with the same properties and weight $ w $.

\begin{theorem}
\label{thm:bounds}
For any fixed $ q, t, k \in \N $, $ q \geq 2 $, the following bounds hold
as $ n \to \infty $:
\begin{align}
\label{eq:bounds}
	\frac{ q^n }{ n^t } \Big(\frac{q}{q-1}\Big)^t
    \, \lesssim \,
      M_q(n; t; k)
    \, \lesssim \,
  \frac{ q^n }{ n^t } \Big(\frac{q}{q-1}\Big)^t  q^{k s} s! (t-s)! ,
\end{align}
where $ s = \big\lfloor \frac{t+1}{q^k+1} \big\rfloor $.
In particular, for $ t = 1 $,
\begin{align}
\label{eq:t1}
   M_q(n; 1; k)  \, \sim \,  \frac{ q^n }{ n } \cdot \frac{q}{q-1} .
\end{align}
\end{theorem}
\begin{IEEEproof}
The lower bound in \eqref{eq:bounds} is a consequence of the construction in
Theorem \ref{thm:construction}.
For fixed $ q, n, w, t, k $, and a Sidon set $ B \subseteq G $ of order $ t $,
the only parameter that is left to be specified in \eqref{eq:construction} is $ b \in G $.
Since the choice of $ b $ can be made in $ |G| $ ways, resulting in at most $ |G| $
(disjoint) codes, and since the total number of $ q $-ary strings of length $ n $
and weight $ w $ is $ \binom{n}{w} (q-1)^w \defeqr S_q(n,w) $, we conclude from
Theorem~\ref{thm:construction} that $ M_q(n, w; t; k)  \geq  S_q(n,w) / |G| $.
By the result of Bose and Chowla \cite{bose+chowla}, the cardinality of the smallest
Abelian group containing a Sidon set of order $ t $ and size $ w $ can be upper
bounded as $ |G| \lesssim w^t $, for any fixed $ t $ and $ w \to \infty $.
This implies that, as $ n \to \infty $ and $ w \sim \omega n $,
\begin{equation}
\label{eq:cwcode}
  M_q(n, w; t; k)  \, \gtrsim \,  \frac{ S_q(n,w) }{ w^t } .
\end{equation}
Now, to obtain the lower bound in \eqref{eq:bounds}, write:
\begin{align}
  M_q(n; t; k)
    \label{eq:aa}
    &=        \sum_{w=0}^n  M_q(n, w; t; k)   \\
    &\geq     \sum_{w = \omega_q n - f(n)}^{\omega_q n + f(n)}  M_q(n, w; t; k)   \\
		\label{eq:1}
		&\gtrsim  \frac{ 1 }{ \big(\omega_q n + f(n)\big)^t }
		           \sum\limits_{w = \omega_q n - f(n)}^{\omega_q n + f(n)}
		            S_q(n,w)   \\
    \label{eq:2}
		&\sim     \frac{ q^n }{ (\omega_q n)^t } ,
\end{align}
where \eqref{eq:aa} holds because the channel does not affect the Hamming
weight of the transmitted string, \eqref{eq:1} follows from \eqref{eq:cwcode},
and \eqref{eq:2} follows from Lemma \ref{thm:typical} and the fact that $ f(n) = o(n) $.

We now turn to the upper bound in \eqref{eq:bounds}.
Let $ \C^* \subseteq \Z_q^n $ be an optimal code correcting $ t $ insertions and
deletions of blocks $ 0^k $, $ |\C^*| = M_q(n; t; k) $.
Consider a codeword $ \myx \in \C^* $ of weight $ w $ and having $ m $ runs of
zeros of length $ \geq k $.
We first observe that the number of strings that can be produced after $ \myx $
is impaired by $ s $ insertions and $ t - s $ deletions of blocks
$ 0^k $ is at least
\begin{equation}
\label{eq:outputs}
  \binom{w + s}{s} \binom{m - s}{t - s} ,
\end{equation}
and that all such strings are of length $ n + k(2s - t) $.
Namely, since $ \weight(\myx) = w $, there are $ w + 1 $ ``bins'' in which blocks
can be inserted, so inserting $ s $ blocks can be done in exactly $ \binom{w + s}{s} $
ways.
On the other hand, deleting $ t - s $ blocks can be done in at least $ \binom{m-s}{t-s} $
ways (we choose $ t - s $ out of $ m $ runs of length $ \geq k $ and delete one
block from each of them; however, we first exclude from these $ m $ runs those runs
into which a block has been inserted in the first step, because otherwise we could
potentially get the same string we started with).
In the asymptotic regime $ n \to \infty $, $ w \sim \omega n $, $ m \sim \mu n $,
the quantity in \eqref{eq:outputs} scales as
\begin{align}
\label{eq:outputs2}
     \sim \,  \binom{\omega n}{s} \binom{\mu n}{t - s}
	\, \sim \,  n^t \frac{\omega^s}{s!} \frac{\mu^{t-s}}{(t-s)!} .
\end{align}
Now, since $ \C^* $ is assumed to correct $ t $ insertions and deletions of blocks
$ 0^k $, the sets of output strings that can be obtained in the above-described
way from any two distinct codewords have to be disjoint.
Since these outputs live in $ \Z_q^{n + k(2s - t)} $, and since, in the asymptotic
regime of interest, we can assume that $ \omega $ and $ \mu $ take on their typical
values $ \omega_q $ and $ \mu_{q,k} $ (see Lemma~\ref{thm:typical}), we conclude that
\begin{align}
\nonumber
  &M_q(n; t; k) \cdot n^t \frac{\omega_q^s}{s!} \frac{\mu_{q,k}^{t-s}}{(t-s)!}
    \, \lesssim \,  q^{n + k(2s - t)}  \\
\label{eq:upper}
  \Leftrightarrow \quad  &M_q(n; t; k)  \, \lesssim \,  \frac{ q^n }{ n^t } \Big(\frac{q}{q-1}\Big)^t  q^{k s} s! (t-s)! .
\end{align}
It is left to optimize the bound over the possible choices of $ s \in \{0, 1, \ldots, t\} $.
To that end note that the sequence $ a_s \defeq q^{k s} s! (t-s)! $ is convex
since $ a_s < \sqrt{ a_{s-1} a_{s+1} } \leq \frac{1}{2} (a_{s-1} + a_{s+1}) $.
This implies that $ a_s $ is minimized at the value of $ s $ for which
$ a_{s} \leq a_{s-1} $ and $ a_{s} < a_{s+1} $.
By checking these conditions directly, we find this value to be
$ s = \big\lfloor \frac{t+1}{q^k+1} \big\rfloor $.
\end{IEEEproof}

\begin{remark}
\textnormal{
Note that the lower bound in \eqref{eq:bounds} is independent of the duplication
length $ k $.
An upper bound independent of $ k $ can also be obtained by choosing a suboptimal
value $ s = 0 $ in \eqref{eq:upper}, which gives
$ M_q(n; t; k) \lesssim \frac{ q^n }{ n^t } \big(\frac{q}{q-1}\big)^t t! $.
Therefore, the dupli\-cation length does not seem to have a significant bearing on
the problem addressed here (see also \eqref{eq:t1}).
\myqed
}
\end{remark}

\begin{remark}[Binary channel with insertions/deletions of zeros]
\label{rem:q2k1}
\textnormal{
Specializing the bounds \eqref{eq:bounds} to $ q = 2 $, $ k = 1 $, we~get:
\begin{align}
\label{eq:bounds21}
	\frac{ 2^n }{ n^t } 2^t
    \, \lesssim \,
      M_2(n; t; 1)
    \, \lesssim \,
  \frac{ 2^n }{ n^t } 2^{t+s} s! (t-s)! ,
\end{align}
where $ s = \big\lfloor \frac{t+1}{3} \big\rfloor $.
The lower bound in \eqref{eq:bounds21} was obtained%
\footnote{Actually, this bound was not stated explicitly in \cite{levenshtein}
because Levenshtein was unaware of the work \cite{bose+chowla} and the construction
of Sidon sets therein.
Consequently, he stated in \cite{levenshtein} an explicit lower bound which is
worse than what his code construction actually implies.}
in \cite[Lem. 3]{levenshtein}.
The upper bound in \eqref{eq:bounds21} strictly improves upon the bound%
\footnote{The upper bound in \cite{levenshtein} is of the same form as the one
in \eqref{eq:bounds21}, but with a suboptimal choice of $ s $: $ s = 0 $ for
$ t $ odd, and $ s = t/2 $ for $ t $ even.}
from \cite[Lem. 2]{levenshtein} for all $ t \geq 3 $.
\myqed
}
\end{remark}

\section*{Acknowledgment}

The authors would like to thank the Associate Editor Marco Baldi and the referees
for their comments which led to a significant improvement in the presentation of
this work.


\begin{thebibliography}{99}

\bibitem{bose+chowla}
   R. C. Bose and S. Chowla,
   ``Theorems in the Additive Theory of Num\-bers,''
   \emph{Comment. Math. Helv.}, vol. 37, no. 1, pp. 141--147, Dec. 1962.
\bibitem{derksen}
   H. Derksen,
   ``Error-Correcting Codes and $ B_h $-Sequences,''
   \emph{IEEE Trans. Inf. Theory}, vol. 50, no. 3, pp. 476--485, Mar. 2004.
\bibitem{dolecek+anantharam}
   L. Dolecek and V. Anantharam,
   ``Repetition Error Correcting Sets: Explicit Constructions and Prefixing Methods,''
   \emph{SIAM J. Discrete Math.}, vol. 23, no. 4, pp. 2120--2146, 2010.
\bibitem{graham+sloane}
   R. L. Graham and N. J. A. Sloane,
   ``Lower Bounds for Constant Weight Codes,''
   \emph{IEEE Trans. Inf. Theory}, vol. 26, no. 1, pp. 37--43, 1980.
\bibitem{jain}
   S. Jain, F. Farnoud, M. Schwartz, and J. Bruck,
   ``Duplication-Correcting Codes for Data Storage in the DNA of Living Organisms,''
   \emph{IEEE Trans. Inf. Theory}, vol. 63, no. 8, pp. 4996--5010, Aug. 2017.
\bibitem{klove}
   T. Kl\o ve,
   ``Error Correcting Codes for the Asymmetric Channel,''
   Technical Report, Dept. of Informatics, University of Bergen, 1981. (Updated in 1995.)
\bibitem{kovacevic+tan}
   M. Kova\v{c}evi\'c and V. Y. F. Tan,
   ``Codes in the Space of Multisets---Coding for Permutation Channels with Impairments,''
   \emph{IEEE Trans. Inf. Theory}, vol. 64, no. 7, pp. 5156--5169, Jul. 2018.
\bibitem{kovacevic}
   M. Kova\v{c}evi\'c,
   ``Runlength-Limited Sequences and Shift-Correcting Codes,''
   preprint \href{https://arxiv.org/abs/1803.06117}{arXiv:1803.06117}, Mar. 2018.
\bibitem{lenz2}
   A. Lenz, N. J\"{u}nger, and A. Wachter-Zeh,
   ``Bounds and Constructions for Multi-Symbol Duplication Error Correcting Codes,''
   preprint \href{https://arxiv.org/abs/1807.02874v1}{arXiv:1807.02874v1}, Jul. 2018.
\bibitem{lenz}
   A. Lenz, A. Wachter-Zeh, and E. Yaakobi,
   ``Duplication-Correcting Codes,''
   \emph{Des. Codes Cryptogr.}, to appear.
	 Published online at: \href{https://doi.org/10.1007/s10623-018-0523-0}{https://doi.org/10.1007/s10623-018-0523-0}.
\bibitem{levenshtein}
   V. I. Levenshtein,
	 ``Binary Codes Correcting Deletions and Insertions of the Symbol $ 1 $'' (in Russian),
	 \emph{Probl. Peredachi Inf.}, vol. 1, no. 1, pp. 12--25, 1965.
\bibitem{mahdavifar+vardy}
   H. Mahdavifar and A. Vardy,
   ``Asymptotically Optimal Sticky-Insertion-Correcting Codes with Efficient Encoding and Decoding,''
   in \emph{Proc. 2017 IEEE Int. Symp. Inf. Theory (ISIT)}, pp. 2683--2687, Aachen, Germany, Jun. 2017.
\bibitem{obryant}
   K. O'Bryant,
   ``A Complete Annotated Bibliography of Work Related to Sidon Sequences,''
   \emph{Electron. J. Combin.}, \#DS11, 39 pp, 2004.
\bibitem{yazdi}
   S. M. H. T. Yazdi, H. M. Kiah, E. Garcia-Ruiz, J. Ma, H. Zhao, and O. Milenkovic,
   ''DNA-Based Storage: Trends and Methods,''
   \emph{IEEE Trans. Mol. Biol. Multi-Scale Commun.}, vol. 1, no. 3, pp. 230--248, Sep. 2015.
\bibitem{yehezkeally+schwartz}
   Y. Yehezkeally and M. Schwartz,
   ``Reconstruction Codes for DNA Sequences with Uniform Tandem-Duplication Errors,''
   preprint \href{https://arxiv.org/abs/1801.06022}{arXiv:1801.06022}, Jan. 2018.
	
\end{thebibliography}
\end{document}